\documentstyle[twocolumn,aps,graphicx]{revtex}
\draft

\begin{document}
\title{Unstable particles versus resonances in impurity systems, \\
conductance in quantum wires}
\author{O.A.~Castro-Alvaredo and A.~Fring}
\address{Institut f\"{u}r Theoretische Physik, Freie Universit\"{a}t
Berlin, Arnimallee 14, D-14195 Berlin, Germany }
\date{\today}
\maketitle

\begin{abstract}
We compute the DC conductance for a homogeneous sine-Gordon model and an
impurity system, which in the conformal limit can be reduced to a Luttinger
liquid, by means of the thermodynamic Bethe ansatz and standard potential
scattering theory. We demonstrate that unstable particles and resonances in
impurity systems lead to a sharp increase of the conductance as a function
of the temperature, which is characterized by the Breit-Wigner formula.
\end{abstract}


\section{Introduction}

In the context of integrable quantum field theories in 1+1 space-time
dimensions a large arsenal of extremely powerful non-perturbative techniques
has been developed over the last two and a half decades. The original
motivation to treat these theories as a testing ground for realistic
theories in higher dimensions is nowadays supplemented by the possibility of
direct applications, since the nanotechnology has advanced to such a degree,
that one dimensional materials, i.e. quantum wires, may be realized
experimentally. A quantity which can be measured directly \cite{Mill} is the
conductance through the quantum wire. There exist also already various
proposals \cite{FLS,LSS} of how to obtain this quantity from general
non-perturbative techniques, such as the thermodynamic Bethe ansatz (TBA) 
\cite{TBAZam,TBAKM} and the form factor approach \cite{Kar} to compute the
current-current two-point correlation functions in the Kubo formula \cite
{Kubo}. Here we want to concentrate on the former approach. Whereas in \cite
{FLS} the emphasis was put on reproducing features of quantum Hall systems
and the authors appealed extensively to massless models, we want to treat
here in contrast systems which are purely massive. In particular we want to
investigate how the properties of unstable particles and impurity resonances
are reflected in a possible conductance measurement.

\section{From conductance to masses of unstable particles}

The direct current $I$ through a quantum wire can be computed simply by
determining the difference of the static charge distributions at the right
and left constriction of the wire, i.e. $I=Q_{R}-Q_{L}$. This is based on
the Landauer transport theory, i.e., on the assumption \cite{FLS,LSS}, that $%
Q(t)\sim (Q_{R}-Q_{L})t\sim $ $(\rho _{R}-\rho _{L})t$, where the $\rho $s
are the corresponding density distribution functions. For more details and a
comparison with the Kubo formula see \cite{CF22}. Placing an impurity in the
middle of the wire, we have to quantify the overall balance of particles of
type $i$ and anti-particles $\bar{\imath}$ carrying opposite charges $%
q_{i}=-q_{\bar{\imath}}\,$\ at the end of the wire at different potentials.
This is achieved once we know the density distribution $\rho _{i}^{r}\left(
\theta ,r,\mu _{i}\right) \,\ $as a function of the rapidity $\theta $, the
inverse temperature $r\,$\ and the chemical potential $\mu _{i}$. In the
described set up half of the particles of one type are already at the same
potential at one of the ends of the wire and the probability for them to
reach the other is determined by the transmission amplitude $\left|
T_{i}\left( \theta \right) \right| $ through the impurity. Therefore 
\begin{eqnarray}
I &=&\sum_{i}I_{i}(r,\mu _{i})  \label{I2} \\
&=&\sum_{i}\int d\theta \frac{q_{i}}{2}\left[ (\rho _{i}^{r}\left( \theta
,r,\mu _{i}^{R}\right) -\rho _{i}^{r}\left( \theta ,r,\mu _{i}^{L}\right)
)\left| T_{i}^{2}\left( \theta \right) \right| \right] .\quad   \nonumber
\label{I22}
\end{eqnarray}
By definition the DC conductance results as 
\begin{equation}
G(r)=\sum\nolimits_{i}G_{i}(r)=\sum\nolimits_{i}\lim_{\mu _{i}\rightarrow
0}I_{i}(r,\mu _{i})\,/\mu _{i}  \label{G}
\end{equation}
and is of course a property of the material itself and a function of the
temperature. In general the expressions in (\ref{I2}) tend to zero for
vanishing chemical potential such that the limit in (\ref{G}) is non-trivial.

Let us now compute the density distribution by means of the thermodynamic
Bethe ansatz. As was pointed out in \cite{Marcio}, the TBA-equations for a
bulk system and a system with a purely transmitting defect are identical.
This is due to the fact that in the thermodynamic limit the number of
defects is kept fixed and is therefore insignificant in thermodynamic
considerations. Therefore the same equations also hold when we allow the
impurity to be such that transmission and reflection are simultaneously
possible. We recall the main equations of the TBA analysis which are
directly relevant in this context, see \cite{TBAZam} for more details and in
particular for the introduction of the chemical potential see \cite{TBAKM}.
For a detailed derivation of the TBA equations in this context see \cite
{CF22}. The main input into the entire analysis is the dynamical interaction
encoded into the scattering matrix $S_{ij}(\theta )$ of two particles of
masses $m_{i}$ and $m_{j}$ and the assumption on the statistical interaction
which we take to be fermionic. As usual \cite{TBAZam,TBAKM}, by taking the
logarithmic derivative of the Bethe ansatz equation and relating the density
of states $\rho _{i}(\theta )$ for particles of type $i$ to the density of
occupied states $\rho _{i}^{r}(\theta )$ one obtains 
\begin{equation}
\rho _{i}(\theta ,r,\mu _{i})=\frac{m_{i}}{2\pi }\cosh \theta
+\sum\nolimits_{j}[\varphi _{ij}\ast \rho _{j}^{r}](\theta )\,.  \label{rho}
\end{equation}
By $\left( f\ast g\right) (\theta )$$:=1/(2\pi )\int d\theta ^{\prime
}f(\theta -\theta ^{\prime })g(\theta ^{\prime })$ we denote the convolution
of two functions and $\varphi _{ij}(\theta )=-id\ln S_{ij}(\theta )/d\theta $%
. The mutual ratio of the densities serves as the definition of the
so-called pseudo-energies $\varepsilon _{i}(\theta )$%
\begin{equation}
\frac{\rho _{i}^{r}(\theta ,r,\mu _{i})}{\rho _{i}(\theta ,r,\mu _{i})}=%
\frac{e^{-\varepsilon _{i}(\theta ,r,\mu _{i})}}{1+e^{-\varepsilon
_{i}(\theta ,r,\mu _{i})}}\,,  \label{dens}
\end{equation}
which have to be positive and real. At thermodynamic equilibrium one obtains
then the TBA-equations, which read in these variables 
\begin{equation}
rm_{i}\cosh \theta =\varepsilon _{i}(\theta ,r,\mu _{i})+r\mu
_{i}+\sum\nolimits_{j}[\varphi _{ij}\ast L_{j}](\theta )\,,  \label{TBA}
\end{equation}
where $r=m/T$, $m_{l}\rightarrow m_{l}/m$, $\mu _{i}\rightarrow \mu _{i}/m$, 
$L_{i}(\theta ,r,\mu _{i})=\ln (1+e^{-\varepsilon _{i}(\theta ,r,\mu _{i})})$%
, with $m$ being the mass of the lightest particle in the model and $T$\ the
temperature. It is important to note that $\mu _{i}$ is restricted to be
smaller than 1. This follows immediately from (\ref{TBA}) by recalling that $%
\varepsilon _{i}\geq 0$ and that for $r$ large $\varepsilon _{i}(\theta
,r,\mu _{i})$ tends to infinity. As pointed out already in \cite{TBAZam},
here just with the small modification of a chemical potential, the
comparison between (\ref{rho}) and (\ref{TBA}) leads to the useful relation 
\begin{equation}
\rho _{i}(\theta ,r,\mu _{i})=\frac{1}{2\pi }\left( \frac{d\varepsilon
_{i}(\theta ,r,\mu _{i})}{dr}+\mu _{i}\right) \,.  \label{rhoe}
\end{equation}
The main task is therefore to solve (\ref{TBA}) for the pseudo-energies from
which then all densities can be reconstructed. In general, due to the
non-linear nature of the TBA-equation, this is done numerically. However, in
the large temperature regime one may carry out various analytical
approximations. For large rapidities and small $r$, one \cite{TBAZam} can
approximate the density of states by 
\begin{equation}
\rho _{i}(\theta ,r,\mu _{i})\sim \frac{m_{i}}{4\pi }e^{|\theta |}\sim \frac{%
1}{2\pi r}\epsilon (\theta )\frac{d\varepsilon _{i}(\theta ,r,\mu _{i})}{%
d\theta }\,,  \label{rr}
\end{equation}
where $\epsilon (\theta )$ is the step function. To obtain this we assume in
(\ref{dens}) that in the large rapidity regime $\rho _{i}^{r}(\theta ,r,\mu
_{i})$ is dominated by (\ref{rr}) and in the small rapidity regime by the
Fermi distribution function, therefore 
\begin{equation}
\rho _{i}^{r}(\theta ,r,\mu _{i})\sim \frac{1}{2\pi r}\epsilon (\theta )%
\frac{d}{d\theta }\ln \left[ 1+\exp (-\varepsilon _{i}(\theta ,r,\mu _{i}))%
\right] \,\,.
\end{equation}
Using this expression, we approximate the current in (\ref{I2}) and for $\mu
_{i}^{R}=-\mu _{i}^{L}=V/2$ the conductance results to 
\begin{eqnarray}
\lim\limits_{r\rightarrow 0}G_{i}(r) &\sim &\frac{q_{i}}{2\pi r}%
\int\nolimits_{-\infty }^{\infty }d\theta \frac{1}{1+\exp [\varepsilon
_{i}(\theta ,r,0)]}  \nonumber \\
&&\times \left. \frac{d\varepsilon _{i}(\theta ,r,V/2)}{dV}\right| _{V=0}%
\frac{d\left[ \epsilon (\theta )\,|T_{i}(\theta )|^{2}\right] }{d\theta }\,.
\label{Ga}
\end{eqnarray}

In order to evaluate (\ref{I2}) and (\ref{Ga}) it remains to specify how to
compute the transmission amplitude. In principle this can be done by
exploiting the factorization equations which result as a consequence of
integrability. However, for systems with a diagonal bulk S-matrix these
equations are not restrictive enough and we will below simply use a free
field expansion and proceed in analogy to standard quantum mechanical
potential scattering. Having obtained $T_{j}(\theta )$ and $R_{j}(\theta )$
one can construct the equivalent quantities for multiple defects from these
functions. Here we are particularly interested in a double defect. Placing
the two defects of the same type at $x_{1}=0$, $x_{2}=y$ the total
transmission amplitude $\hat{T}_{j}$ can be build up from the ones of a
single defect as \cite{DMS} 
\begin{equation}
\hat{T}_{j}(\theta )=\frac{T_{j}^{2}(\theta )}{1-R_{j}^{2}(\theta )\exp
(i2y\sinh \theta )}\,\,.  \label{double}
\end{equation}

Having assembled all the ingredients for the computation of $G$ we turn to
the question of how the properties of unstable particles are reflected in
this quantity? Assuming that $S_{ij}(\theta )$ possesses a resonance pole at 
$\theta _{R}=\sigma -i\bar{\sigma}$, the Breit-Wigner formula \cite{BW}
allows to determine the mass $M_{\tilde{c}}$ and the decay width $\Gamma _{%
\tilde{c}}$ of an unstable particle of type $\tilde{c}$%
\begin{eqnarray}
2M_{\tilde{c}}^{2} &=&\sqrt{\gamma ^{2}+\tilde{\gamma}^{2}}+\gamma \geq
2(m_{i}+m_{j})^{2}\quad  \label{BrW} \\
\Gamma _{\tilde{c}}^{2}/2 &=&\sqrt{\gamma ^{2}+\tilde{\gamma}^{2}}-\gamma
\geq 4m_{i}m_{j}(1-\cosh \sigma \cos \bar{\sigma}),  \label{BrW2}
\end{eqnarray}
where $\gamma =m_{i}^{2}{}+m_{j}^{2}{}+2m_{i}m_{j}\cosh \sigma \cos \bar{%
\sigma}$ and $\tilde{\gamma}=2m_{i}m_{j}\sinh |\sigma |\sin \bar{\sigma}\,$.
The thresholds in (\ref{BrW}) and (\ref{BrW2}) result from energetic reasons 
\cite{CF4}. We will now demonstrate that besides unstable particles also
resonances in impurity systems can be described by means of the Breit-Wigner
formula. An important consequence of (\ref{BrW}) is that we can approximate
the mass in there by $M_{\tilde{c}}^{2}\approx 1/2m_{i}m_{j}(1+\cos \bar{%
\sigma})\exp |\sigma |$ for large $\sigma $. Then under a renormalization
group flow $M_{\tilde{c}}\rightarrow r_{C}M_{\tilde{c}}$ the quantity $M_{%
\tilde{c}}\sim r_{C}^{1}e^{\sigma _{1}/2}=r_{C}^{2}e^{\sigma _{2}/2}$
remains invariant. Once the unstable particle can be created, it can
participate in the overall conductance and one should observe an increase at 
$T_{C}$ in $G$ related to this process. For this interpretation to hold, we
should observe the following scaling behaviour of the conductance 
\begin{equation}
G(r_{C}^{1},\sigma _{1})=G(r_{C}^{2},\sigma _{2})\,\,\,\quad \text{for\quad }%
\,r_{C}^{1}e^{\sigma _{1}/2}=r_{C}^{2}e^{\sigma _{2}/2}.  \label{M12}
\end{equation}
Surely $r_{C}$ is not sharply defined, but taking for instance the middle
between the beginning and the end of the onset seems reasonably well
identifiable. Here $r_{C}$ is the inverse of the critical temperature $%
r_{C}=m/T_{C}$ at which the unstable particle for fixed $\sigma $ is formed.
This means the identification of the onset in a conductance measurement will
provide $r_{C}$, such that for given $\sigma $ the mass of the unstable
particle can be deduced.

\section{Homogeneous sine-Gordon model}

The $SU(3)_{2}$-homogeneous sine-Gordon (HSG) model is the simplest of its
kind and contains only two self-conjugate solitons, which we denote by
``+'', ``$-$'', and one unstable particle, which we call $\tilde{c}$. The
corresponding scattering matrix was found \cite{HSGS} to be $S_{\pm \pm }=-1$%
, $S_{\pm \mp }(\theta )=\pm \tanh \left( \theta \pm \sigma -i\pi /2\right)
/2$, which means the resonance pole is situated at $\theta _{R}=\mp \sigma
-i\pi /2$. Stable bound states may not be formed. It is known \cite{DMS},
that integrable parity invariant impurity systems with a diagonal bulk
S-matrix, apart from $S=\pm 1$, do not allow simultaneously non-trivial
reflection and transmission amplitudes. This statement can be extended to
the parity violating case \cite{CF20}. We treat therefore (\ref{I2}) for a
transparent defect, i.e. $|T|=1$. The results for the conductance after
solving numerically the TBA equations for $\mu _{R}=-\mu _{L}=0.25$ are
depicted in figure 1.

\begin{center}
\includegraphics[width=8.2cm,height=6.09cm]{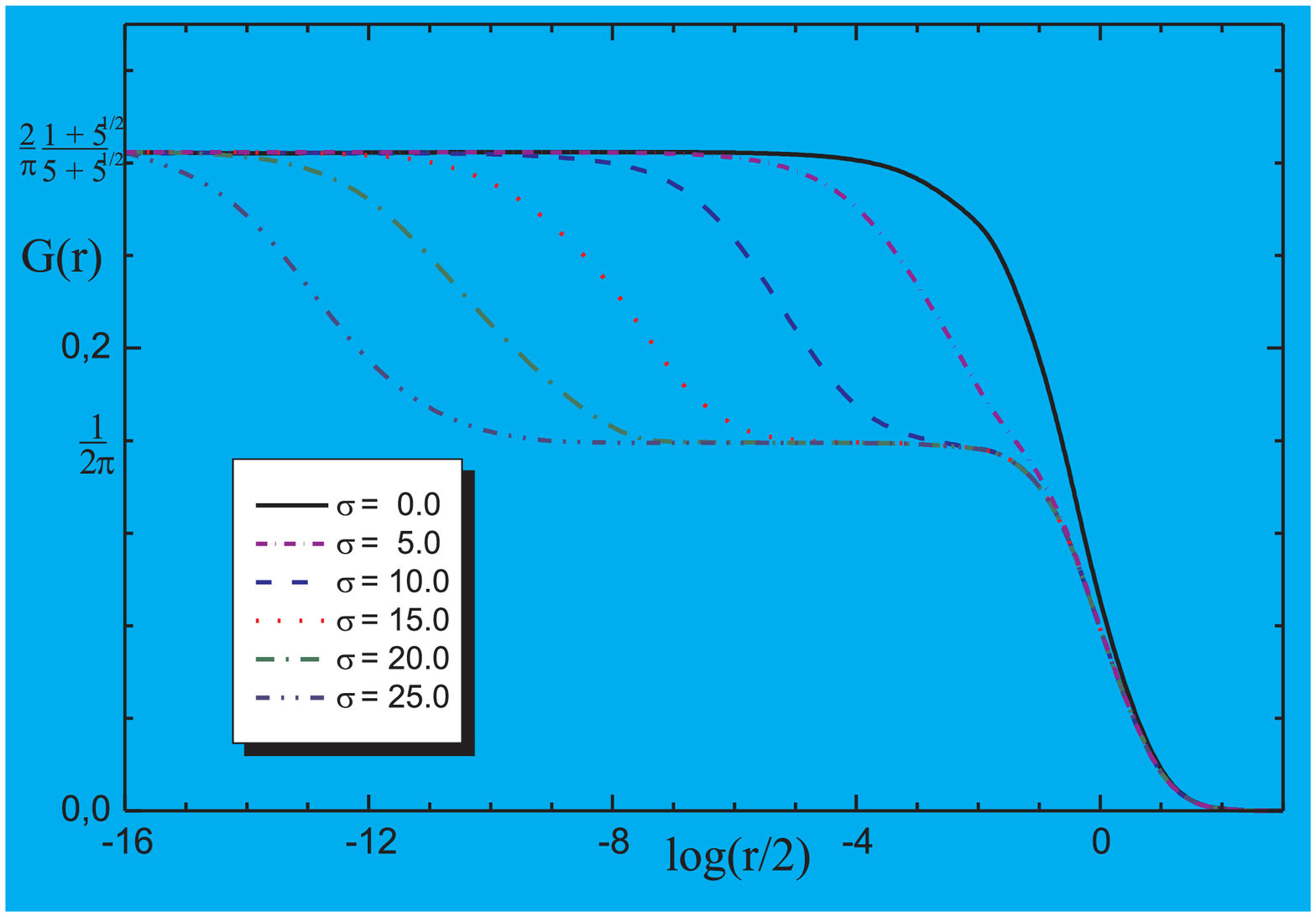}
\end{center}

\vspace*{0.2cm}

\noindent {\small FIG 1: Conductance }$G${\small \ for the $SU(3)_{2}$%
-HSG-model as a function of }$\log (r/2)=\log (m/2T)${\small \ for various
values of the resonance parameter }$\sigma ${\small .}

To carry out the limit $\mu \rightarrow 0$ is rather complicated when one
does not have an explicit analytic expression at hand as in our case.
However, we can take the result for finite $\mu$ as a very good
approximation, since we observe that $G(r)/ \mu \sim const$ for small $r$.
We observe the onset of the unstable particle in form of a relatively sharp
increase in $G$ and in particular the validity of (\ref{M12}). The
interpretation is clear: Only when we reach an energy scale at which the
unstable particle can be formed it can participate in the conducting
process. All this information is encoded in the density $\rho
_{i}^{r}(\theta ,r,\mu _{i})$. Also the bound in (\ref{BrW}) is respected.
Computing now $\varepsilon _{i}(\theta ,0,0)$ in a standard TBA fashion,
e.g., \cite{CFKM}, we predict analytically the plateaux from (\ref{Ga}) at $%
2(1+\sqrt{5})/(5+\sqrt{5})\pi $ and at $1/2\pi $. The latter value is
obtained from the fact that in the region in which $\sigma \gg -2 \log(r/2)$%
, the system can be viewed as consisting out of two free Fermions such that (%
\ref{Ga}) gives the quoted value.

\section{Free Fermion with impurities}

The continuous version of the 1+1 dimensional Ising model with a line of
defect was first treated in \cite{Cabra}. Thereafter it has also been
considered in \cite{GZ,DMS} and \cite{Konik} from a different point of view.
In \cite{Cabra,GZ,DMS} the impurity was taken to be of the form of the
energy operator and in \cite{Konik} also a perturbation in form of a single
Fermion has been considered. Here we also include a further type of defect.

Let us consider the Lagrangian density for a complex free Fermion $\psi $
with $\ell $ defects 
\begin{equation}
{\cal L}=\bar{\psi}(i\gamma ^{\mu }\partial _{\mu }-m)\psi
\,+\sum\nolimits_{n=1}^{\ell }\delta (x-x_{n}){\cal D}_{n}(\bar{\psi},\psi
)\,,  \label{La}
\end{equation}
where we describe the defect by the functions ${\cal D}_{n}(\bar{\psi},\psi
),$ which we assume to be linear in the Fermi fields. In the following we
will restrict ourselves to the case $\ell =2$ with $x_{n}=ny$ and ${\cal D}%
_{n}(\bar{\psi},\psi )={\cal D}(\bar{\psi},\psi )$. We compute the
transmission amplitude as indicated in \cite{GZ,DMS,Konik}, namely by
decomposing the solution to these equations as $\psi (x)=\Theta (x)$ $\psi
_{+}(x)+\Theta (-x)$ $\psi _{-}(x)$ and substituting them into the equations
of motion. This way we obtain the constraints 
\begin{equation}
i\gamma ^{1}(\psi _{+}(x)-\psi _{-}(x))|_{x=x_{n}}=\left. \frac{\partial 
{\cal D}_{n}(\bar{\psi},\psi )}{\partial \bar{\psi}}\right| _{x=x_{n}}\,\,.
\label{bcon}
\end{equation}
Using now the standard Fourier expansion for a complex free Fermi field, the
transmission and reflection amplitudes can be read off componentwise from (%
\ref{bcon}) as the coefficients of $a_{j,-}^{\dagger }(\theta )=R_{\bar{%
\jmath}}(\theta )a_{j,-}^{\dagger }(-\theta )$, $a_{j,-}^{\dagger }(\theta
)=T_{\bar{\jmath}}(\theta )a_{j,+}^{\dagger }(\theta )$, etc.

Recalling now that for the free Fermion the TBA-equations are simply solved
by $\varepsilon _{i}(\theta ,r,\mu _{i})=rm_{i}\cosh \theta -r\mu _{i}\,$,\
we compute 
\begin{equation}
G(r)=\frac{r}{2\pi }\int\nolimits_{0}^{\infty }d\theta \frac{\cosh \theta
\,\left| T_{i}\left( \theta \right) \right| ^{2}}{1+\cosh (r\cosh \theta )}%
\,.  \label{gg}
\end{equation}
To proceed further we have to specify the impurity.

\subsection{The energy operator defect, ${\cal D}(\bar{\protect\psi},\protect%
\psi )=g\bar{\protect\psi}\protect\psi $}

\noindent From (\ref{bcon}) we compute 
\begin{eqnarray}
R_{j}(\theta ,B) &=&R_{\bar{\jmath}}(\theta ,B)=-\frac{i\sin B\cosh \theta }{%
\sinh \theta +i\sin B}\,,  \label{t1} \\
T_{j}(\theta ,B) &=&T_{\bar{\jmath}}(\theta ,B)=\frac{\cos B\sinh \theta }{%
\sinh \theta +i\sin B}\,,  \label{t2}
\end{eqnarray}
where we used a common parameterization in this context $\sin
B=-4g/(4+g^{2}) $. The expressions $R_{\bar{\jmath}}(\theta ,B)$ and $T_{%
\bar{\jmath}}(\theta ,B)$ coincide with the solutions found in \cite{DMS},
which, however, in general does not correspond to taking our particles
simply to be self-conjugate, since we use Dirac Fermions. Using (\ref{t1})
and (\ref{t2}) we compute with (\ref{double}) the conductance for a double
defect with varying distance $y$. The results of our numerical computations
for the conductance are depicted in figure 2.

In the high temperature regime we can confirm once more these data by some
analytical computations. From (\ref{Ga}) we obtain 
\begin{equation}
\frac{1}{2\pi }\left( \frac{\cos ^{2}B}{1+\sin ^{2}B}\right) ^{2}\leq G(r,y)
< \frac{1}{2\pi }\,\,.  \label{g4}
\end{equation}
The lower bound becomes exact for $y/r<1$. For $B=0.51$ the values $0.0602$
are well reproduced in figure 2.

\begin{center}
\includegraphics[width=8.2cm,height=6.09cm]{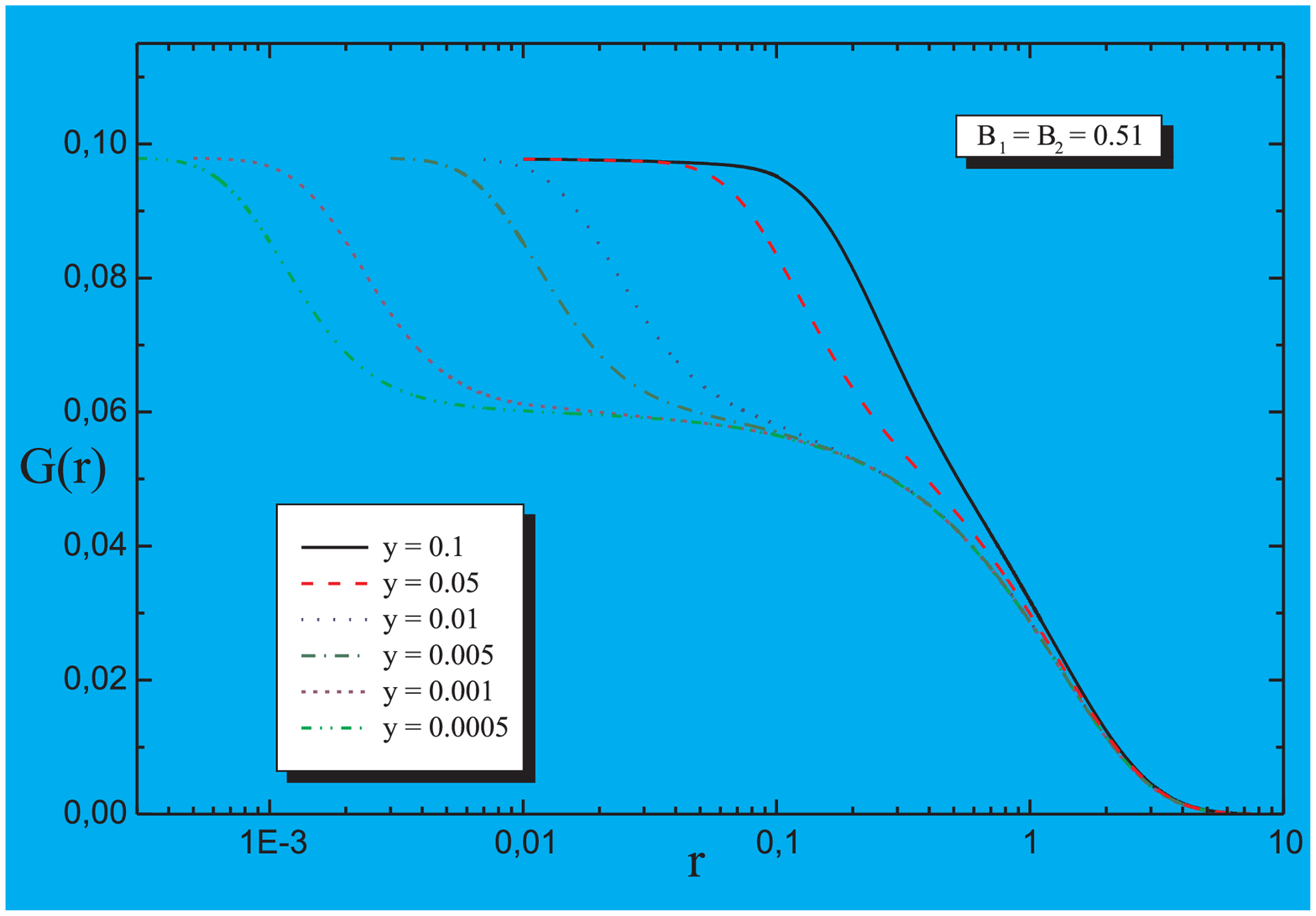}
\end{center}

\noindent {\small FIG 2: Conductance }$G${\small \ for the free Fermion with
double energy defect at distances }$y${\small \ as a function of the inverse
temperature }$r=m/T${\small .}

We observe a similar type of behaviour as in the preceding section and
denote again the point of onset in the conductance by $r_{C}$. Then, we
deduce from our data that the following scaling relations

\begin{equation}
G(r_{C}^{1},y_{1})=G(r_{C}^{2},y_{2})\quad \text{for \quad }%
r_{C}^{2}y_{1}=r_{C}^{1}y_{2}\,.  \label{==}
\end{equation}
Comparison with (\ref{M12}) suggests that we can relate the distance between
the two defects to the resonance parameter as $\sigma =2 \ln ($const$/y)$.
However, despite the fact that the net result with regard to the conductance
is the same, the origin of the onset is different. Whereas in the previous
section it resulted from a change in the density distribution function it is
now triggered by the structure of $|\hat{T}(\theta )|$. Since $\rho ^{r}$
keeps its overall shape and just moves its peak with varying temperature,
the onset has to occur when $|\hat{T}(\theta )|$ reaches its maxima. Using (%
\ref{t1}), (\ref{t2}) and (\ref{double}), it is easy to verify that $| \hat{T%
}(\theta =\ln [(2n+1)\pi /y]) | \approx 1$. Drawing now an analogy to the
scattering matrix, this value plays the same role as $\theta _{R}$ and we
therefore identify 
\begin{equation}
\sigma_n =\ln [(2n+1)\pi /y]\,.  \label{sig}
\end{equation}
Having fixed the resonance parameter $\sigma $ we may, in view of (\ref{==}%
), relate the temperature to the mass scale of the unstable particle,
associated now to the resonance, analogously as in the discussion after (\ref
{M12}). However, there are some differences. Whereas in the HSG-model the
onset is attributed to a single particle, the effect for the double defect
system is attributed to several resonances. We identify $\sigma \approx
\sigma_0 + \sigma_1$. The other difference is that $y$ is now a measurable
quantity, such that $\sigma $ in (\ref{sig}) can be experimentally
determined. On the other hand the sigma in (\ref{M12}) is usually a free
parameter in the HSG-type models. Let us now verify our observations for a
different type of defect.

\subsection{Luttinger type liquid, ${\cal D}(\bar{\protect\psi},\protect\psi
)=\bar{\protect\psi}(g_{1}+g_{2}\protect\gamma ^{0})\protect\psi $}

There exist various ways to realize Luttinger type liquids \cite{Lutt}.
Taking the conformal limit of the defect ${\cal D}(\bar{\psi},\psi )=\bar{%
\psi}(g_{1}+g_{2}\gamma ^{0})\psi $, we obtain an impurity which played a
role in this context \cite{Aff} when setting the bosonic number counting in
there to be one. Analogously to the previous sections we compute the related
transmission and reflection amplitudes 
\begin{eqnarray*}
R_{j}(\theta ,g_{1},g_{2}) &=&\frac{4i(g_{2}-g_{1}\cosh \theta )}{%
4i(g_{1}-g_{2}\cosh \theta )+(4+g_{1}^{2}-g_{2}^{2})\sinh \theta }\,, \\
T_{j}(\theta ,g_{1},g_{2}) &=&\frac{(4+g_{2}^{2}-g_{1}^{2})\sinh \theta }{%
4i(g_{1}-g_{2}\cosh \theta )+(4+g_{1}^{2}-g_{2}^{2})\sinh \theta }\,.
\end{eqnarray*}
The expressions for the particle $\bar{\jmath}$ are obtained by replacing $%
g_{1}\rightarrow -g_{1}$. The results of our numerical computation for $%
g_{1}=0.7$ and $g_{2}=0.2$ depicted in figure 3 confirm the same physical
picture as outlined in the previous subsection. Our analytical prediction
for the lowest plateau from (\ref{Ga}) is $0.0324$.

\begin{center}
\includegraphics[width=8.2cm,height=6.09cm]{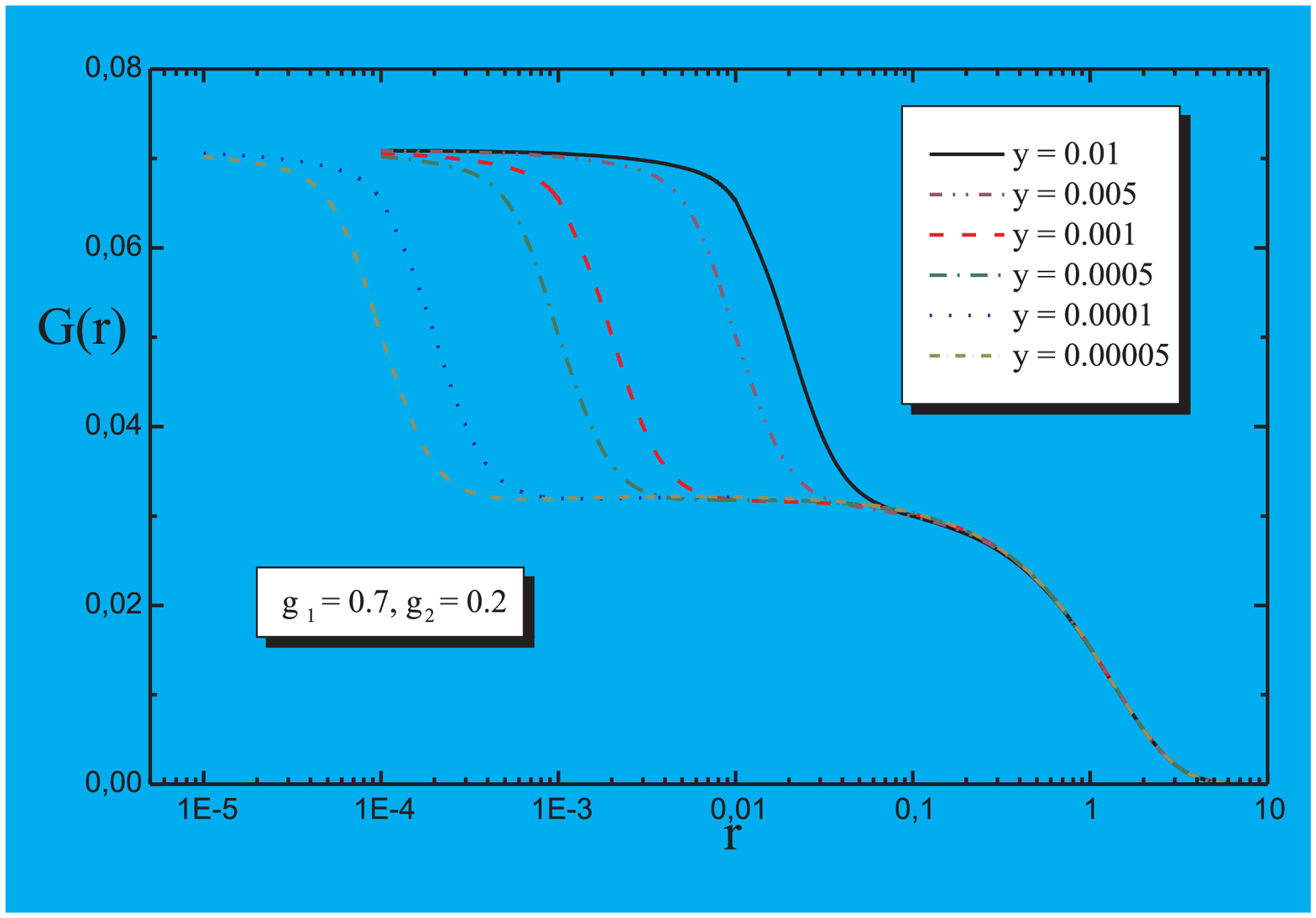}
\end{center}

\vspace*{0.2cm}

\noindent {\small FIG 3: Conductance }$G${\small \ for the free Fermion with
two defects $D_{n}(\bar{\psi},\psi )=\bar{\psi}(g_{1}+g_{2}\gamma ^{0})\psi $
at distance }$y${\small \ as a function of the inverse temperature }$m/T$%
{\small .}

\section{Conclusions}

By using the TBA to compute the density distribution function and
relativistic potential scattering theory to determine the transmission
amplitude, we evaluated the DC conductance by means of equation (\ref{I2}).
We demonstrated that the sharp increase of the conductance as a function of
the temperature can be attributed to the presence of unstable particles in
the HSG models or likewise to a resonance of a double defect system.\medskip

{\bf Acknowledgments:} We are grateful to the Deutsche
Forschungsgemeinschaft (Sfb288) for financial support. We thank F.
G\"{o}hmann for valuable comments.

\end{document}